\documentstyle[PASJadd,psfig]{PASJ95}
\draft
\newcommand{\vol}{}
\newcommand{\no}{}
\newcommand{\lett}{}
\newcommand{\spage}{}
\newcommand{\rdate}{1997 August 27}
\newcommand{\adate}{1997 November 21}

\begin{document}

\title{Dispersion of Inferred SNe Ia/SNe II Ratios for Current Models
of Supernova Nucleosynthesis}

\author{Shigehiro {\sc Nagataki},$^1$ Masa-aki {\sc Hashimoto},$^2$,
and Katsuhiko {\sc Sato}$^{1,3}$
\\[12pt]
$^1$ {\it Department of Physics, School of Science, the University
of Tokyo, 7-3-1 Hongo, Bunkyoku, Tokyo 113 }\\
{\it E-mail(TY): Nagataki@utaphp1.phys.s.u-tokyo.ac.jp}\\
$^2$ {\it Department of Physics, Faculty of Science,
Kyusyu University, Ropponmatsu, Fukuoka 810} \\
$^3$ {\it Research Center for the Early Universe, School of
Science, the University of Tokyo, 7-3-1 Hongo, Bunkyoku, Tokyo 113}
}

\abst{
We have estimated the dispersion of the inferred relative frequencies
of Type-Ia and Type-
II supernovae ($N_{\rm Ia}/ N_{\rm II}$) in our Galaxy by fitting the
numerical results of supernova nucleosynthesis to the solar-system
abundances. The ratio $N_{\rm Ia}/ N_{\rm II}$ is estimated to be
0.056--0.14, which is consistent with the observation, if the model of
Woosley and Weaver (1995, WW95) is adopted for Type-II supernovae (SNe II). 
On the other hand, the upper limit of $N_{\rm Ia}/ N_{\rm II}$
becomes too large for the model of Hashimoto (1995, Ha95).
However, Ha95 can fit the solar values better than WW95 as far as the
abundant nuclei are concerned.
These results mean that Ha95 can reproduce the main
nuclei of the solar-system well and that WW95 can reproduce the solar
values over a wide mass-number range.
We also note that $N_{\rm Ia}/ N_{\rm II}$ tends to become smaller
if the delayed detonation model is adopted for
Type-Ia supernovae (SNe Ia).
The dispersion of $N_{\rm Ia}/ N_{\rm II}$ obtained in the present
investigation is
larger than that concluded by Tsujimoto et al.\ (1995), which will 
have an influence on the estimate of 
the average life time of SNe Ia's progenitors and/or the
star-formation rate history in our galaxy. 
}

\kword{Chemical evolution --- Solar system --- Sun: abundances --- Supernovae}

\maketitle
\thispagestyle{headings}

\section
{Introduction}

Supernovae play an important role in ejecting heavy elements of $A$
$\ge$ 12 produced during stellar evolution and/or an explosion.
There are two types of supernovae: Type-I and Type-II 
supernovae. Type-Ia supernovae (SNe Ia) are thought
to be thermonuclear explosions of accreting white dwarfs (e.g.,
Nomoto et al.\ 1994). On the other hand, Type-II supernovae (SNe II)
and Type-Ib/Ic supernovae (SNe Ib/Ic) are thought to be
collapse-driven supernovae (e.g., Bethe 1990).
Many calculations of them have been performed and their chemical
compositions have been calculated.

As for SNe Ia, the defraglation model has been thought to be more
favored than the detonation model, since $\rm ^{56}Ni$ is over-produced
in the detonation model.  
However, concerning the amount of $\rm ^{56}Ni$, the defraglation model can
explain the spectrum of SNe Ia fairly well (e.g., Nomoto et al.\ 1997a).
In particular, a model W7 (Nomoto et al.\ 1984) has been
considered to be 'standard', and the calculated chemical composition has been
used widely to explain the solar-system abundances.
As for SNe II, two groups have systematically calculated the nucleosynthesis
(e.g., 
Woosley, Weaver 1995; Hashimoto 1995). Although their calculations
assumed different physical process such as the criterion of
convection, the $\rm
^{12}C(\alpha,\gamma)^{16}O$ rate, which is crucial for constructing
stellar models, and initial shock-wave conditions, which determines the
peak temperature during the explosion,  
they succeeded in explaining the chemical compositions of SNe II
with a fair degree of precision.
In fact, the solar-system abundances are well reproduced within a factor of
2--3 by the combination of their results for a wide range of mass numbers,
except for some nuclei (Hashimoto 1995; Timmes et al.\ 1995.)

By using the results of W7 and Ha95, the relative frequencies 
of SNe Ia and SNe II in our galaxy was obtained to be $N_{\rm
Ia}/N_{\rm II} =
0.15$ (Tsujimoto et al. 1995), which is consistent with the fact that the 
observed estimate of the SNe Ia frequency is about $10 \%$ of the total
supernovae occurrence (van den Bergh, Tammann 1991).
However, there is one question in their calculations. They concluded
that the contribution of SNe Ia to the solar-system abundances is in
the range $r = 0.09 \pm 0.01$ against possible uncertainties in their
analysis (see subsection 2.2).
However, they arbitrarily selected 31 species to be fitted, and
neglected other nuclei when they derived the proper value of $r$. 
We think that it is worth examining whether the dispersion of the $N_{\rm
Ia}/N_{\rm II}$ changes or not by the choice of the nuclei to be fitted.
Moreover, we should examine whether the proper value of $r$ can
be really predicted from numerical calculations without such an
arbitrariness.

Another motivation for the present study is the model dependence
of the prediction of $N_{\rm Ia}/N_{\rm II}$.
Recently, based on the progress of
numerical calculations for supernovae, the chemical compositions of
their ejecta have been recalculated.
In particular, the effect of the delayed detonation has been taken into
consideration for SNe Ia.
As for the W7 model, the flame speed is assumed to be relatively high, 
which is inconsistent with that of multi-dimensional calculations;
the simple defraglation can not generate an energetic
explosion, or can not lead to an explosion in the multi-dimensional
calculations (Arnett, Livne 1994; Khokhlov 1995). 
To solve this problem, a mechanism of delayed detonation is
proposed by Khokhlov (1991). 
Recently, nucleosynthesis in SNe Ia was recalculated based on
the assumption of delayed detonation (Nomoto et al.\ 1997b), and has presented
a more excellent explanation for the observation of SNe Ia. For example,
the velocity profiles of intermediate nuclei, such as Si and S,
become wider compared with that of the W7 model, which is consistent with the
observations. 
We also note that the chemical compositions of SNe II
are different between Ha95 and WW95 (Hashimoto 1995; Woosley, Weaver 1995).

To investigate the capability of reproducing the solar-system abundances
with using the avairable results, we will explore the model
dependence and derive the dispersion based on numerical calculations. 
It is important to estimate the dispersion, since the change in
the relative frequency of SNe Ia and SNe II will have an
influence on the estimate of the average lifetime of SNe Ia's
progenitors and/or the star-formation history in our galaxy.

We describe our calculation method
in section 2. The results are presented in section
3. A summary and a discussion are given in section 4.

\section{ Models and Method of Calculations } \label{calculation}

\subsection{Models of SNe Ia and SNe II} \label{SNeIa}
\indent
As for the composition of SNe Ia, we use the result of W7 for 
the simple deflagration model and WDD2 for the
delayed detonation model, respectively (Nomoto et al.\ 1984; Nomoto et 
al.\ 1997b).
We also use the results of Ha95 and WW95 for 
SNe II. More precisely, we use for Ha95 the results of helium-star models with 
masses of
$M_{\alpha}=3.3, 4, 5, 6, 8, 16,$ and $32$, which approximately correspond
to main-sequence stars of $M_{\rm ms}$ = 13, 15, 18,
20, 25, 40, and 70 $M_{\odot}$, and for WW95 S13A,
S15A, S18A, S20A, S25A, and S40A models.
The averaged synthesized yields in 10--50$M_{\odot}$ stars are obtained with a
weight of Salpeter's initial mass function, as follows:
\begin{eqnarray}
M_{i,\rm II}=\frac{\int M_{i}(m)m^{-(1+x)} dm}{\int m^{-(1+x)} dm}.
\end{eqnarray}
The differences in the calculated chemical compositions are shown in figure 1.
The open circles denote W7/WDD2 and the dots represent WW95/Ha95.
The abscissa represents the mass number. The plotted 46 nuclei,
which are the same 
nuclei as those in table 1 of Tsujimoto et al. (1995), are
summarized in table 1.


\subsection{Calculation Method for the Solar-System Abundances} \label{solar}
\indent

In an analysis of the reproduction of the solar-system abundances, we 
adopt the same method with Tsujimoto et al.\ (1995). 
At first,  as the sum of $M_{i,\rm Ia}$ we define $M_{\rm Ia}$ which are the
total heavy element mass of SNe Ia. We also define $M_{\rm II}$ in the
same way.

Next, we define the abundance pattern $x_i$ as
\begin{eqnarray}
x_i = rM_{i,\rm Ia}/M_{\rm Ia} + (1-r)M_{i,\rm II}/M_{\rm II}  \;\;\;
(0 \le r \le 1), 
\end{eqnarray}
which is to be compared with the solar $x_i(\odot)$,
defined as $Z_i/\sum_i Z_i$, where $Z_i$ is
the observed abundance of the $i$-th element per unit mass (Anders,
Grevesse 1989).
The most probable value of $r=r_{\rm p}$ is
determined by minimizing the following function (Yanagida et al.\ 1990):

\begin{eqnarray}
g(r) = \sum_{i=1}^{n} [\log x_i - \log x_i(\odot)]^2/n,
\end{eqnarray}
where $i$ runs over the heavy elements and their isotopes chosen in
the minimization procedure. We summarize those elements in table 1.
We note that $r$ has a different meaning from that of the relative
frequency $N_{\rm Ia}/N_{\rm II}$. The relation between them is
\begin{eqnarray}
\frac{N_{\rm Ia}}{N_{\rm II}} = \frac{\omega_{\rm II}}{\omega_{\rm Ia}}
\frac{M_{\rm II}}{M_{\rm Ia}}\frac{r_{\rm p}}{(1-r_{\rm p})},
\end{eqnarray}
where $\omega_{\rm Ia}$ and $\omega_{\rm II}$ represent the mass fraction of
heavy elements ejected into the interstellar gas from SNe Ia and SNe
II, respectively. These values were estimated to be 0.27 and 0.22 in
the solar neighborhood from the numerical calculation (Tsujimoto et al.\ 1995).
We calculated $g(r)$ in five cases. We used all 46
nuclei (Case 1). Next, we used the selected 31 nuclei (Case 2) and 
the most abundant 6 nuclei (Case 3).
Then, 20 elements, each of which is a summation of the isotopic abundances,
were examined (Case 4). Finally, the selected 14 elements were fitted (Case 5).
Furthermore, we also performed the same analysis for LMC and SMC with 
the same elements with Tsujimoto et al.\ (1995) to check our conclusion.
The value of $M_{\rm II}/M_{\rm Ia}$ for each case is summarized in table 2.
We note that Tsujimoto et al.\ (1995) concluded that $r_{\rm p} = 0.09
\pm 0.01$ and $N_{\rm Ia}/N_{\rm II} = 0.15$ in their analysis by adopting
Case 2 and Case 5 for the results of W7+Ha95.


\section{ Results} \label{results}

\subsection{The Solar System Abundances Pattern }
\indent

We show in figure 2 the form of $g(r)$ for four models: 
W7+Ha95, W7+WW95, WDD2+Ha95, and WDD2+WW95.
The thick solid, long-dashed, short-long-dashed, short-
dashed, and dot lines correspond to Case 1, 2, 3, 4, and 5,
respectively.

Generally speaking, $g(r)$ has a minimum value at $r \sim 0.1$ in most
cases. In particular, $g(r)$ has
minimum values for all cases if WW95 is adopted.
On the other hand,  $g(r)$ is not sensitive
to $r$ for the W7+Ha95 and WDD2+Ha95 models in Case 1 and 4.
However, $g(r)$ has a deep minimum value in the case of 3 if Ha95 is
adopted. These results mean that WW95 can reproduce the solar values 
well in many nuclei and that Ha95 can reproduce them well
for selected abundant nuclei.
The values of $r_{\rm p}$ for all cases are summarized in table 3. 
We should note that the range of $r_{\rm p}$ is wider than that of Tsujimoto et
al.\ (1995).

We can see the effect of delayed detonation.
The values of $r_{\rm p}$ tend to become smaller if WDD2 is adopted.
We also feel that the minimum values of $g(r_{\rm p})$ seem to become smaller
in WDD2. This means that the solar-system abundances are fitted
better by WDD2 than W7.
We examine this tendency by adopting our analysis 
to SMC and LMC in the next subsection.


\subsection{The SMC and LMC Abundance Patterns }

In subsection 3.1, we found that both $r_{\rm p}$ and $g(r_{\rm p})$
become smaller
if WDD2 is adopted. We will check this tendency by
adopting our method of analysis to other galaxies: SMC 
and LMC. As was done by Tsujimoto et al. (1995),12 and 10 elements were
taken into consideration for SMC and
LMC, respectively (see table 1). The results are given in table 4.
We found the
same tendency of $r_{\rm p}$ and $g(r_{\rm p})$, which supports our
conclusion. We also note the $r_{\rm p}$s obtained in this analysis tend to
be larger than those of the solar-system abundances, which is consistent
with Tsujimoto et al. (1995), even if WDD2 is adopted.


\subsection{Relative frequency of SNe Ia/SNe II}

The ratio $N_{\rm Ia}$/$N_{\rm II}$ depends on the three values of
$\omega_{\rm II}$/$\omega_{\rm
Ia}$, $M_{\rm II}$/$M_{\rm Ia}$, and $r_{\rm p}$. In our analysis,
$\omega_{\rm II}$/$\omega_{\rm Ia}$ is constant (subsection 2.2). WW95
tend to make $M_{\rm II}$/$M_{\rm Ia}$ smaller than Ha95.
WDD2 also make it lower, though its degree is weak. 
The range of $N_{\rm Ia}$/$N_{\rm II}$ in table 3 is estimated from 
$r_{\rm p}$ and $M_{\rm II}$/$M_{\rm Ia}$ for Case 1--5.
The values in the parenthesis in table 3 are not so accurate, since $g(r)$ is
insensitive to $r$.
We note that $N_{\rm Ia}/ N_{\rm II}$ is estimated to be
0.056--0.14, which would be consistent with the observation
if the model of WW95 is adopted. 
On the other hand, the upper limit of $N_{\rm Ia}/ N_{\rm II}$
becomes too large for Ha95.
We also note that $N_{\rm Ia}/ N_{\rm II}$ tends to become smaller
if WDD2 is adopted.
Generally speaking, the range of $N_{\rm Ia}$/$N_{\rm II}$ is fairly
wider compared with the conclusion of Tsujimoto et al.\ (1995). 

\subsection{ Other Analysis}

The method described in subsection 2.2 is not the only one that can be 
used to determine $r_{\rm p}$.
In this subsection we examine another method to obtain $r_{\rm p}$:
$\chi ^2$ fitting.
We use two elements, O and Fe, for the $\chi ^2$ fitting.
The $r_{\rm p}$ is determined to be $7.5 \times
10^{-2}$--$1.0 \times 10^{-1}$, $(6.8$--$9.1) \times 10^{-2}$,
$(1.1$--$1.3) \times 10^{-1}$, and $9.8 \times
10^{-2}$--$1.2 \times 10^{-1}$ (99$\%$ C.L.) for W7+Ha95, WDD2+Ha95, W7+WW95,
and WDD2+WW95, respectively. 
However, we must say that all models are
excluded unless proper nuclei are chosen for the fitting, since 
the uncertainty of the observation is much smaller than that of the
theoretical prediction. That is why these ranges mentioned above are less
persuasive than those discussed in subsection 3.1.
In other words,
we must wait for progress concerning theoretical calculations
to use this method effectively.

\subsection{Present Situation of Numerical Simulations}

In figure 3,
we show the solar abundance pattern $x_i/x_i({\odot})$ (see, equation (2))
predicted by numerical
calculations using $r_{\rm p}$ of Case 2.
The open circles represent WW95 and the dots denote Ha95.
There are some nuclei which are less produced in the range of mass numbers
35--50. For example, Ha95 has a less-production problem of some
elements, such as Ar, K, Ti.
WW95 also has a problem of less production of $\rm ^{41}K$, $\rm ^{44}Ca$,
and Ti. Cu and Zn are also less produced by Ha95.
On the other hand, some
nuclei in the range 50--65 are over produced. These problems remaine to be
solved to predict $r_{\rm p}$ more precisely. 

\section{ Summary and Discussion} \label{summary}
\indent

We have explored the dispersion of $N_{\rm Ia}/ N_{\rm II}$
using the four models Ha95, WW95, W7, and WDD2.
As for $r_{\rm p}$, $g(r)$ clearly has minimum values if WW95 is adopted.
On the other hand, $g(r)$ is insensitive
to $r$ for the W7+Ha95 and WDD2+Ha95 models in Case 1 and 4.
However, $g(r)$ has a deep minimum value in the case of 3 if Ha95 is
adopted. These results mean that WW95 reproduces the solar values well in many 
nuclei and that Ha95 reproduces them well especially with regard to
abundant nuclei.
Among the most abundant 6 nuclei, the
abundance of $\rm ^{56}Fe$ is different most between WW95 and Ha95.
This would be because only WW95 determines the mass cut by
a hydrodynamical calculation. This suggests the difficulty 
of the determining the mass cut from the hydrodynamics, even if the
spherical explosion is assumed.
$M_{\rm II}$/$M_{\rm Ia}$ tends to become lower for WW95 than for Ha95.
Although WDD2 also make it lower, its degree is weak.

As a result, the range of $N_{\rm Ia}$/$N_{\rm II}$ becomes
0.056--0.14 for WW95+SNe Ia
, which is consistent with the observation, and
0.077--(0.61) for Ha95+SNe Ia. 
We also note that $N_{\rm Ia}/ N_{\rm II}$ tends to become smaller
if the delayed detonation model is adopted for
Type-Ia supernovae.
Since the delayed-detonation model may be a more realistic model than
the simple defraglation model, this tendency should be stressed.
Generally speaking, the dispersion of $N_{\rm Ia}/ N_{\rm II}$ obtained
in our study is larger than that concluded to be $\sim 10 \%$
by Tsujimoto et al.\ (1995), as can be seen in table 3.
We think that this is the present situation of
the predictability of numerical calculations, and that this
dispersion must always be taken into account.
For example, our results will bring the dispersion of both
the average lifetime of SNe Ia's progenitors and history of the
star-formation rate in our galaxy. Moreover, 
the critical mass, which represents the upper-mass limit of
SNe II progenitor, may also be changed.

A realistic s-process calculation will increase the amount of some
less-produced
elements in Ha95, as is shown by Prantzos et al. (1990).
It would hold true for WW95.
This suggests that the neutron-capture process during stellar evolution
is also important for some elements. Since WW95 used a larger network for the
'post processing', they obtained 
neutron-rich elements. On the other hand, Ha95 used
a small network during stellar evolution.
As a result, neutron-rich nucleosynthesis seems 
to be insufficient.
We also note that there is a possibility that
the chemical composition
of SNe Ia could be changed if the mass-accretion rate is changed
(Nomoto et al.\ 1997b). An asymmetric explosion in SNe II may also
help to solve
these problems,
since it would produce more $\rm ^{44}Ca$ and Ti than the spherical calculations
(Nagataki et al.\ 1997a). The initial conditions of the shock wave should also
be considered (Nagataki et al.\ 1997b) because they affect the production of
the iron-peak elements.

Finally, let us to say that there remains still some
physical uncertainties to be studied in order to clarify the
uncertainty of supernova nucleosynthesis, which could be greater than
the dispersion obtained in the present study.
Therefore, we must trace the physical uncertainty in order to find the
reason for the
over/under production problems.
Such efforts
will cause the dispersion of $N_{\rm Ia}/ N_{\rm II}$ to be smaller, and we
should then be able to discuss the chemical evolution in the universe more precisely.

\par
\vspace{1pc}\par
We are grateful to an anonymous referee for useful comments.
We also thank for Ken'ichi Nomoto for useful discussions.
This research has been
supported in part by a Grant-in-Aid for the Center-of-Excellence (COE) 
Research (07CE2002) and for the Scientific Research Fund (05243103,
07640386, 3730) of the Ministry of Education, Science, Sports and
Culture in Japan
and by Japan Society for the Promotion of Science Postdoctoral
Fellowships for Research Abroad.

\clearpage
\section*{References}
\re
Anders E., Grevesse N.\ 1989, Geochim. Cosmochin. Acta 53, 197
\re
Arnett W.D., Livne E.\ 1994, ApJ 427, 315
\re
Bethe H.A.\ 1990, Rev. Mod. Phys.\ 62, 801
\re
Hashimoto M.\ 1995, Prog. Theor. Phys.\ 94, 663 
\re
Khokhlov A.M.\ 1991, A\&A 246, 383
\re
Khokhlov A.M.\ 1995, ApJ 449, 695
\re
Nagataki S., Hashimoto M., Sato K., Yamada S.\ 1997a, ApJ 486, 1026
\re
Nagataki S., Hashimoto M., Yamada S.\ 1997b, PASJ 49,
\re
Nomoto K., Iwamoto K., Kishimoto N.\ 1997a, Science 276, 1378
\re
Nomoto K., Thielemenn F.-K., Yokoi K.\ 1984, ApJ 286, 644
\re
Nomoto K., Yamaoka H.,
Shigeyama T., Kumagai S., Tsujimoto T.\ 1994, in Supernovae (Les 
Houches, Session LIV), ed J.\ Audouze, S.\ Bludman, R.\ Mochkovitch,
J. Zinn-Justin (Elsevies Sci. Publ.,
Amsterdam) p199
\re
Nomoto, K. et al.\ 1997b, Nucl. Phys. A, in press.
\re
Prantzos N., Hashimoto M., Nomoto K. 1990, A\&A 234, 211
\re
Timmes F.X., Woosley S.E., Weaver T.A.\ 1995, ApJS 98, 617
\re
Tsujimoto T., Nomoto K., Yoshii Y., Hashimoto M., Yanagida S.,
Thielemann F.-K.\ 1995, MNRAS 277, 945
\re
van den Bergh S., Tammann G.\ 1991, ARA\&A 29, 363
\re
Woosley S.E., Weaver T.A.\ 1995, ApJS 101, 181 
\re
Yanagida S., Nomoto K., Hayakawa S.\ 1990, in Proc. 21st International
Cosmic Ray Conference 4, 44
\re
\clearpage
\begin{table*}
\small
\begin{center}
Table~1. \hspace{4pt} Nuclei used in the analysis.\\
\end{center}
\vspace{3pt}
\renewcommand{\arraystretch}{0.5}
\begin{tabular*}{175mm}{c|c|c|c|c|c|c|c|cccccc}
\hline \hline
Species & Case 1 & Case 2 & Case 3 & Case 4 & Case 5 & LMC & SMC & W7 & WDD2 & Ha95 & WW95 \\[4pt]\hline \\[-10pt]
$\rm ^{16}O$  & $\bigcirc$ & $\bigcirc$ & $\bigcirc$ & $\bigcirc$ & $\bigcirc$ & $\bigcirc$ & $\bigcirc$ & 1.43E-01 & 6.93E-02 & 1.80E-00 & 1.14E-00\\
$\rm ^{18}O$  & $\bigcirc$ & $\bigcirc$ &            & $\bigcirc$ & $\bigcirc$ & $\bigcirc$ & $\bigcirc$ & 8.25E-10 & 4.62E-07 & 4.61E-03 & 4.04E-03\\
$\rm ^{20}Ne$ & $\bigcirc$ & $\bigcirc$ & $\bigcirc$ & $\bigcirc$ & $\bigcirc$ & $\bigcirc$ & $\bigcirc$ & 2.02E-03 & 9.13E-04 & 2.12E-01 & 1.65E-01\\
$\rm ^{21}Ne$ & $\bigcirc$ & $\bigcirc$ &            & $\bigcirc$ & $\bigcirc$ & $\bigcirc$ & $\bigcirc$ & 8.46E-06 & 1.47E-06 & 1.08E-03 & 6.07E-04\\
$\rm ^{22}Ne$ & $\bigcirc$ & $\bigcirc$ &            & $\bigcirc$ & $\bigcirc$ & $\bigcirc$ & $\bigcirc$ & 2.49E-03 & 1.96E-06 & 1.83E-02 & 1.88E-02\\
$\rm ^{23}Na$ & $\bigcirc$ & $\bigcirc$ &            & $\bigcirc$ & $\bigcirc$ &            & $\bigcirc$ & 6.32E-05 & 1.30E-05 & 6.51E-03 & 5.13E-03\\
$\rm ^{24}Mg$ & $\bigcirc$ & $\bigcirc$ & $\bigcirc$ & $\bigcirc$ & $\bigcirc$ & $\bigcirc$ & $\bigcirc$ & 8.50E-03 & 4.76E-03 & 8.83E-02 & 3.83E-02\\
$\rm ^{25}Mg$ & $\bigcirc$ & $\bigcirc$ &            & $\bigcirc$ & $\bigcirc$ & $\bigcirc$ & $\bigcirc$ & 4.05E-05 & 2.39E-05 & 1.44E-02 & 9.15E-03\\
$\rm ^{26}Mg$ & $\bigcirc$ & $\bigcirc$ &            & $\bigcirc$ & $\bigcirc$ & $\bigcirc$ & $\bigcirc$ & 3.18E-05 & 3.57E-05 & 2.01E-02 & 1.18E-02\\
$\rm ^{27}Al$ & $\bigcirc$ & $\bigcirc$ &            & $\bigcirc$ & $\bigcirc$ &            & 	      & 9.86E-04 & 2.74E-04 & 1.48E-02 & 7.96E-03\\
$\rm ^{28}Si$ & $\bigcirc$ & $\bigcirc$ & $\bigcirc$ & $\bigcirc$ & $\bigcirc$ &            & $\bigcirc$ & 1.50E-01 & 2.71E-01 & 1.05E-01 & 1.14E-01\\
$\rm ^{29}Si$ & $\bigcirc$ & $\bigcirc$ &            & $\bigcirc$ & $\bigcirc$ &            & $\bigcirc$ & 8.61E-04 & 3.87E-04 & 8.99E-03 & 3.61E-03\\
$\rm ^{30}Si$ & $\bigcirc$ & $\bigcirc$ &            & $\bigcirc$ & $\bigcirc$ &            & $\bigcirc$ & 1.74E-03 & 6.35E-04 & 8.05E-03 & 4.02E-03\\
$\rm ^{31}P$  & $\bigcirc$ & $\bigcirc$ &            & $\bigcirc$ & $\bigcirc$ &            & 	     & 4.18E-04 & 1.80E-04 & 1.21E-03 & 1.14E-03\\
$\rm ^{32}S$  & $\bigcirc$ & $\bigcirc$ & $\bigcirc$ & $\bigcirc$ & $\bigcirc$ & $\bigcirc$ & $\bigcirc$ & 8.41E-02 & 1.65E-01 & 3.84E-02 & 5.34E-02\\
$\rm ^{33}S$  & $\bigcirc$ & $\bigcirc$ &            & $\bigcirc$ & $\bigcirc$ & $\bigcirc$ & $\bigcirc$ & 4.50E-04 & 2.49E-04 & 1.78E-04 & 3.18E-04\\
$\rm ^{34}S$  & $\bigcirc$ & $\bigcirc$ &            & $\bigcirc$ & $\bigcirc$ & $\bigcirc$ & $\bigcirc$ & 1.90E-03 & 2.50E-03 & 2.62E-03 & 4.33E-03\\
$\rm ^{35}Cl$ & $\bigcirc$ &            & 	     & $\bigcirc$ &            & 	     & & 1.34E-04 & 9.83E-05 & 1.01E-04 & 3.80E-04\\
$\rm ^{37}CL$ & $\bigcirc$ &            &            & $\bigcirc$ &            & 	     & & 3.98E-05 & 3.36E-05 & 1.88E-05 & 7.59E-05\\
$\rm ^{36}Ar$ & $\bigcirc$ & $\bigcirc$ &            & $\bigcirc$ & $\bigcirc$ & $\bigcirc$ & $\bigcirc$ & 1.49E-02 & 3.35E-02 & 6.62E-03 & 9.05E-03\\
$\rm ^{38}Ar$ & $\bigcirc$ & $\bigcirc$ &            & $\bigcirc$ & $\bigcirc$ & $\bigcirc$ & $\bigcirc$ & 1.06E-03 & 1.45E-03 & 1.37E-03 & 2.22E-03\\
$\rm ^{39}K$  & $\bigcirc$ &            &            & $\bigcirc$ & &            &  & 8.52E-05 & 9.00E-05 & 6.23E-05 & 2.38E-04\\
$\rm ^{41}K$  & $\bigcirc$ &            &            & $\bigcirc$ &            &            & & 7.44E-06 & 7.12E-06 & 5.07E-06 & 5.31E-06\\
$\rm ^{40}Ca$ & $\bigcirc$ & $\bigcirc$ &            & $\bigcirc$ & $\bigcirc$ & $\bigcirc$ & $\bigcirc$ & 1.23E-02 & 3.45E-02 & 5.77E-03 & 6.53E-03\\
$\rm ^{44}Ca$ & $\bigcirc$ &            &            & $\bigcirc$ & $\bigcirc$ & $\bigcirc$ & $\bigcirc$ & 8.86E-06 & 2.07E-05 & 5.53E-05 & 2.49E-05\\
$\rm ^{46}Ti$ & $\bigcirc$ &            &            & $\bigcirc$ &            &            & & 1.71E-05 & 1.76E-05 & 7.48E-06 & 2.60E-05\\
$\rm ^{47}Ti$ & $\bigcirc$ &            &            & $\bigcirc$ &            &            & & 6.04E-07 & 1.24E-06 & 2.11E-06 & 4.94E-06\\
$\rm ^{48}Ti$ & $\bigcirc$ &            &            & $\bigcirc$ &            &            & & 2.03E-04 & 8.53E-04 & 1.16E-04 & 3.17E-05\\
$\rm ^{49}Ti$ & $\bigcirc$ &            &            & $\bigcirc$ &            &            & & 1.69E-05 & 6.71E-05 & 5.98E-06 & 4.10E-06\\
$\rm ^{50}Ti$ & $\bigcirc$ &            &            & $\bigcirc$ &            &            & & 1.26E-05 & 4.60E-04 & 3.81E-10 & 6.39E-06\\
$\rm ^{50}Cr$ & $\bigcirc$ & $\bigcirc$ &            & $\bigcirc$ & $\bigcirc$ & $\bigcirc$ & $\bigcirc$ & 2.71E-04 & 5.24E-04 & 4.64E-05 & 8.16E-05\\
$\rm ^{52}Cr$ & $\bigcirc$ & $\bigcirc$ &            & $\bigcirc$ & $\bigcirc$ & $\bigcirc$ & $\bigcirc$ & 5.15E-03 & 2.01E-02 & 1.15E-03 & 2.32E-04\\
$\rm ^{53}Cr$ & $\bigcirc$ & $\bigcirc$ &            & $\bigcirc$ & $\bigcirc$ & $\bigcirc$ & $\bigcirc$ & 7.85E-04 & 2.26E-03 & 1.19E-04 & 2.52E-05\\
$\rm ^{54}Cr$ & $\bigcirc$ & $\bigcirc$ &            & $\bigcirc$ & $\bigcirc$ & $\bigcirc$ & $\bigcirc$ & 1.90E-04 & 2.03E-03 & 2.33E-08 & 1.54E-05\\
$\rm ^{55}Mn$ & $\bigcirc$ & $\bigcirc$ &            & $\bigcirc$ & $\bigcirc$ & $\bigcirc$ & $\bigcirc$ & 8.23E-03 & 1.88E-02 & 3.86E-04 & 1.94E-04\\
$\rm ^{54}Fe$ & $\bigcirc$ & $\bigcirc$ &            & $\bigcirc$ & $\bigcirc$ & $\bigcirc$ & $\bigcirc$ & 1.04E-01 & 7.08E-02 & 3.62E-03 & 6.52E-03\\
$\rm ^{56}Fe$ & $\bigcirc$ & $\bigcirc$ & $\bigcirc$ & $\bigcirc$ & $\bigcirc$ & $\bigcirc$ & $\bigcirc$ & 6.13E-01 & 6.15E-01 & 8.44E-02 & 1.66E-02\\
$\rm ^{57}Fe$ & $\bigcirc$ & $\bigcirc$ &            & $\bigcirc$ & $\bigcirc$ & $\bigcirc$ & $\bigcirc$ & 2.55E-02 & 1.39E-02 & 2.72E-03 & 5.90E-04\\
$\rm ^{59}Co$ & $\bigcirc$ &            &            & $\bigcirc$ &            &            & & 1.02E-03 & 8.60E-04 & 7.27E-05 & 1.43E-04\\
$\rm ^{58}Ni$ & $\bigcirc$ & $\bigcirc$ &            & $\bigcirc$ & $\bigcirc$ & $\bigcirc$ & $\bigcirc$ & 1.28E-01 & 3.34E-02 & 3.63E-03 & 7.93E-03\\
$\rm ^{60}Ni$ & $\bigcirc$ & $\bigcirc$ &            & $\bigcirc$ & $\bigcirc$ & $\bigcirc$ & $\bigcirc$ & 1.05E-02 & 4.15E-03 & 1.75E-03 & 9.18E-04\\
$\rm ^{62}Ni$ & $\bigcirc$ & $\bigcirc$ &            & $\bigcirc$ & $\bigcirc$ & $\bigcirc$ & $\bigcirc$ & 2.66E-03 & 1.36E-03 & 5.09E-04 & 2.94E-04\\
$\rm ^{63}Cu$ & $\bigcirc$ &            &            & $\bigcirc$ &             &            & & 1.79E-06 & 4.25E-05 & 8.37E-07 & 2.61E-05\\
$\rm ^{65}Cu$ & $\bigcirc$ &            &            & $\bigcirc$ &             &            & & 6.83E-07 & 1.59E-05 & 4.07E-07 & 7.47E-05\\
$\rm ^{64}Zn$ & $\bigcirc$ &            &            & $\bigcirc$ &             &            & & 1.22E-05 & 6.97E-07 & 1.03E-05 & 3.93E-05\\
$\rm ^{66}Zn$ & $\bigcirc$ &            &            & $\bigcirc$ &             &            & & 2.12E-05 & 3.18E-05 & 8.61E-06 & 1.23E-04\\
\hline
\end{tabular*}
\vspace{6pt}\par\noindent
$\bigcirc$ Nuclei used in the analysis of $g(r)$.
\end{table*}

\begin{table*}
\begin{center}
Table~2. \hspace{4pt} $M_{\rm II}$/$M_{\rm Ia}$ for each model.\\
\end{center}
\vspace{6pt}
\begin{tabular*}{\textwidth}{@{\hspace{\tabcolsep}
\extracolsep{\fill}}p{6pc}ccccccl}
\hline \hline\\[-6pt]
Case        & W7+Ha95 & W7+WW95 & WDD2+Ha95 & WDD2+WW95 \\[4pt]\hline\\[-6pt]
46 nuclei   & 1.85    & 1.23    &  1.79  & 1.19 \\ 
31 nuclei   & 1.86    & 1.24    &  1.80  & 1.20 \\
 6 nuclei   & 2.33    & 1.53    &  2.07  & 1.36 \\
20 elements & 1.85    & 1.23    &  1.79  & 1.19 \\
14 elements & 1.86    & 1.24    &  1.80  & 1.20 \\
\hline
\end{tabular*}
\end{table*}

\begin{table*}
\begin{center}
Table~3. \hspace{4pt} The most proper value of $r_{\rm p}$.\\
\end{center}
\vspace{6pt}
\begin{tabular*}{\textwidth}{@{\hspace{\tabcolsep}
\extracolsep{\fill}}p{6pc}cccccl}
\hline \hline\\[-6pt]
Case        & W7+Ha95  & W7+WW95 & WDD2+Ha95 & WDD2+WW95 \\[4pt]\hline \\[-6pt]
46 nuclei   & (2.8E-01)& 1.1E-01 & (1.4E-01) & 6.1E-02   \\ 
31 nuclei   & 9.2E-02  & 9.2E-02 & 5.0E-02   & 5.4E-02   \\
 6 nuclei   & 8.5E-02  & 1.0E-01 & 7.6E-02   & 9.2E-02   \\
20 elements & (2.9E-01)& 1.1E-01 & (1.8E-01) & 7.7E-02   \\
14 elements & 8.1E-02  & 9.6E-02 & 6.4E-02   & 6.3E-02   \\
\hline 
$N_{\rm Ia}/ N_{\rm II}$ & 1.3E-01--(6.1E-01) & 1.0E-01--1.4E-01&
7.7E-02--(3.3E-01) & 5.6E-02--1.1E-01 \\
\hline
\end{tabular*}
\vspace{6pt}\par\noindent
Parenthesis represents that $g(r)$ is insensitive to $r$. In that
case, $r_{\rm p}$ is hard to be determined.
\end{table*}

\begin{table*}
\begin{center}
Table~4. \hspace{4pt} The most proper value of $r_{\rm p}$/$g(r_{\rm p})$ for SMC
and LMC.\\
\end{center}
\vspace{6pt}
\begin{tabular*}{\textwidth}{@{\hspace{\tabcolsep}
\extracolsep{\fill}}p{6pc}cccl}
\hline \hline\\[-6pt]
Object & W7+Ha95 & W7+WW95 & WDD2+Ha95 & WDD2+WW95 \\[4pt]\hline \\[-6pt]
LMC    & 1.6E-01/5.2E-02 & 1.6E-01/6.5E-02 & 1.1E-01/3.6E-02   &
9.6E-02/4.7E-02    \\ 
SMC    & 1.9E-01/4.7E-02 & 1.9E-01/4.1E-02 & 1.5E-01/3.7E-02   &
1.2E-01/2.9E-02   \\ 
\hline
\end{tabular*}
\end{table*}

\clearpage
\centerline{Figure Captions}
\bigskip
\begin{fv}{1}
{7cm}
{Comparison of chemical compositions. The open circles represent WDD2/W7
and the dots denote WW95/Ha95}
\end{fv}
\begin{fv}{2}
{7cm}
{$g(r)$ for each model. The thick-solid, long-dashed, short-long-dashed, short-
dashed, and dot lines correspond to Case 1, 2, 3, 4, and 5,
respectively.
}
\end{fv}
\begin{fv}{3}
{7cm}
{Comparison with solar values. The open circles represent Ha95+SNe Ia and
the dots denote WW95+SNe Ia. Upper: W7 is used for SNe Ia. Lower: WDD2 is used.
}
\end{fv}

\thispagestyle{empty}
\begin{figure}
\begin{center}
   \leavevmode\psfig{figure=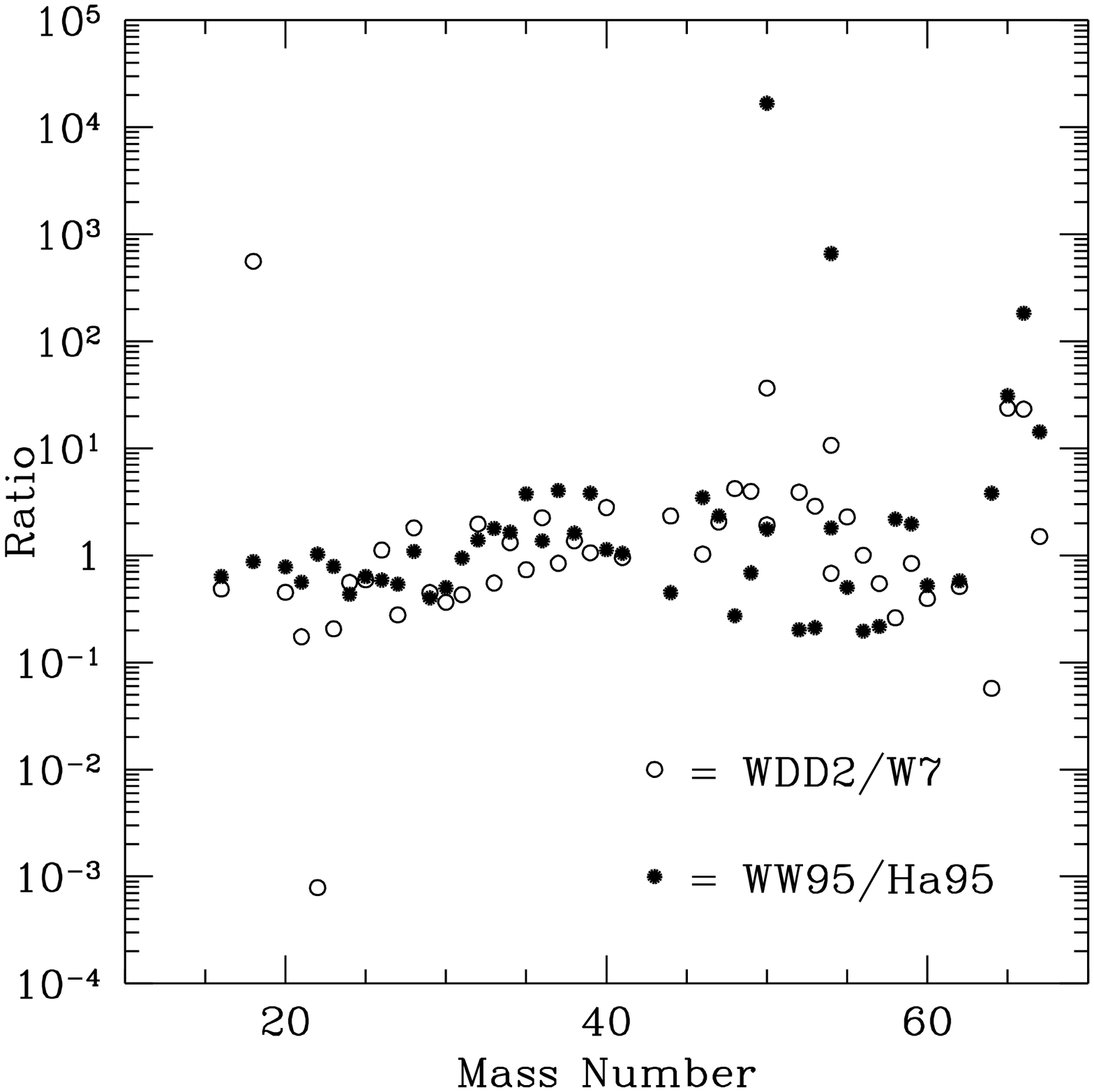,height=7cm,angle=0}
\end{center}
\caption{}
\label{diff1}
\end{figure}

\thispagestyle{empty}
\begin{figure}
\begin{center}
   \leavevmode\psfig{figure=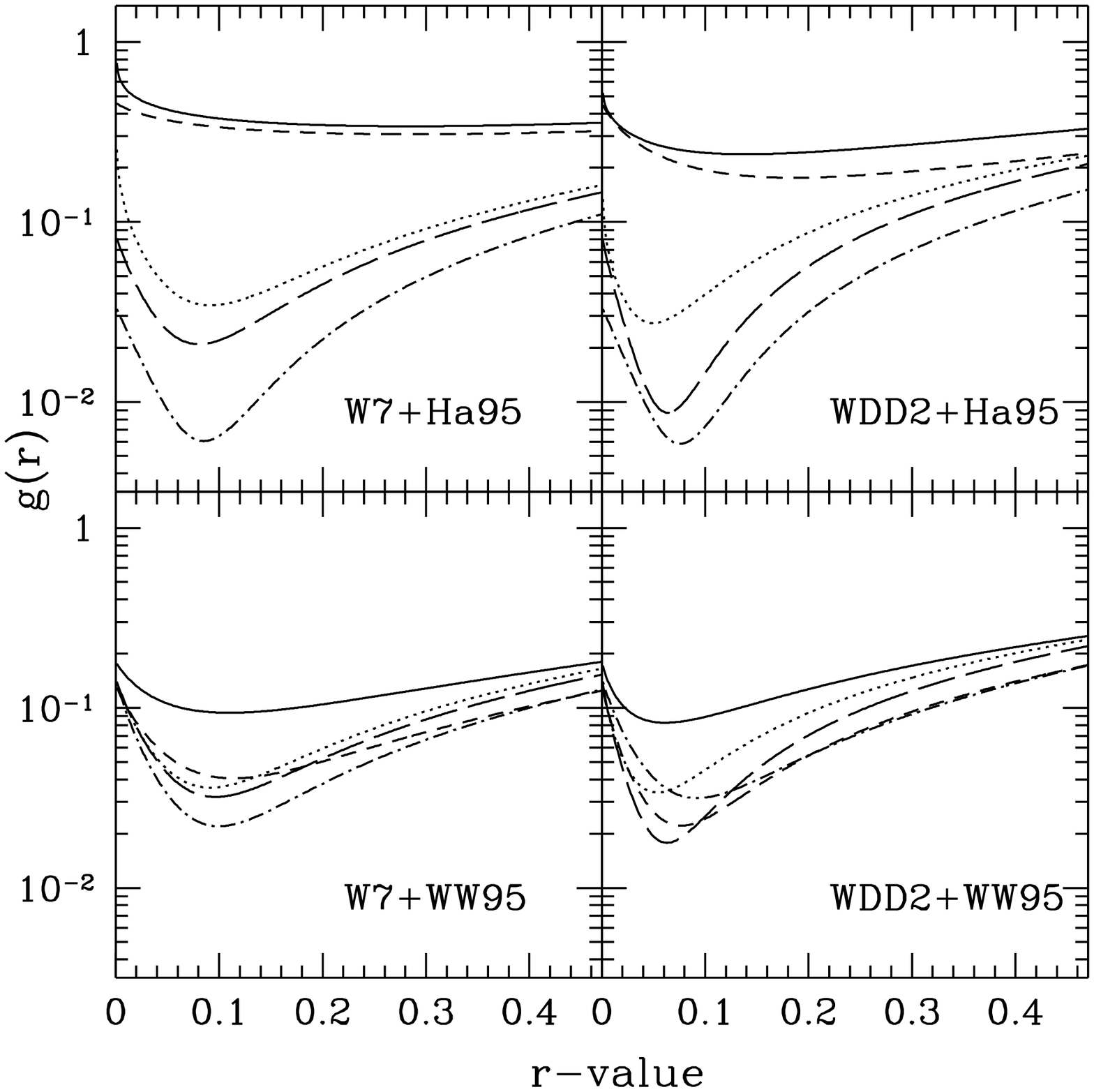,height=7cm,angle=0}
\end{center}
\caption{}
\label{diff2}
\end{figure}

\thispagestyle{empty}
\begin{figure}
\begin{center}
   \leavevmode\psfig{figure=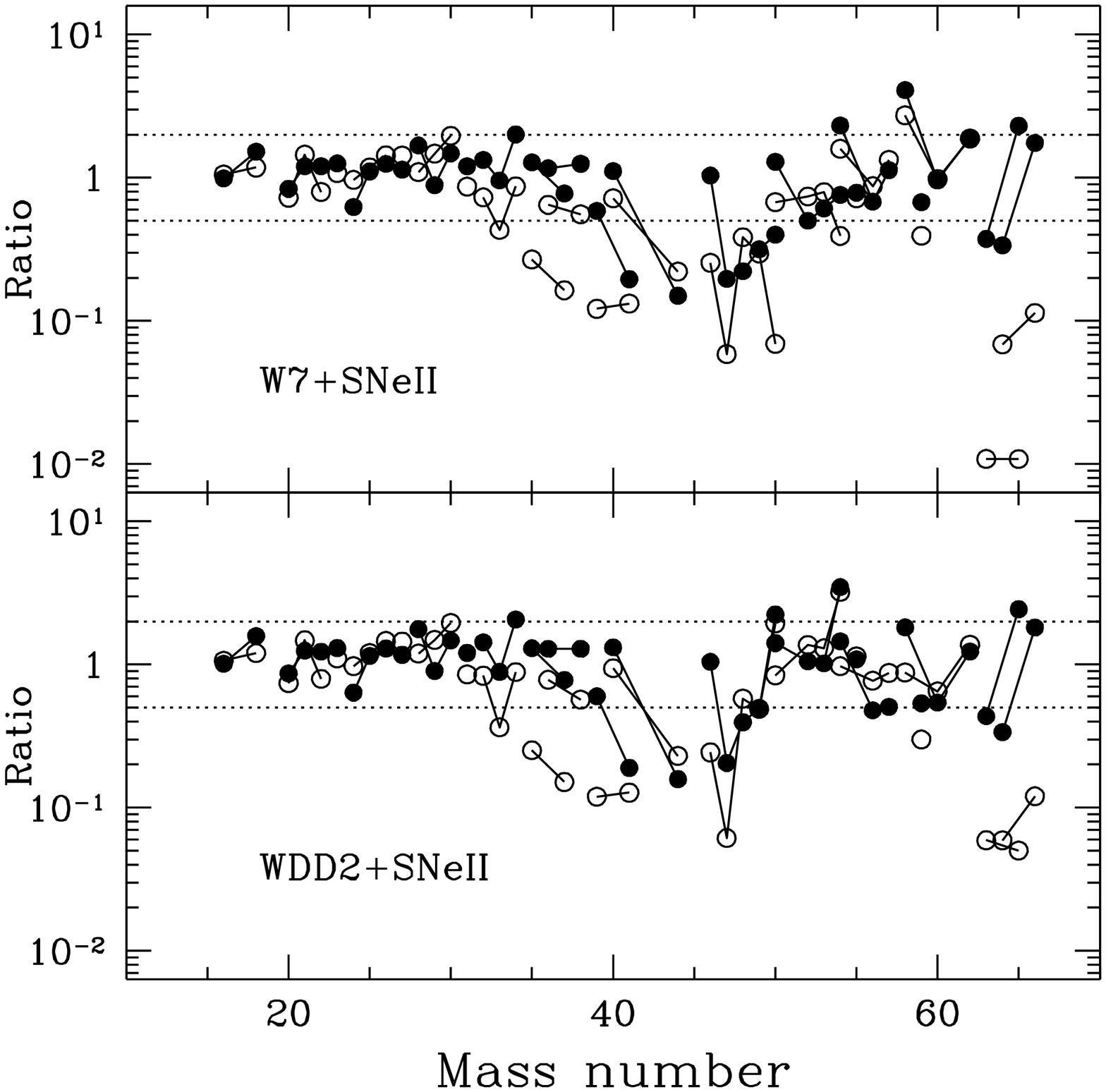,height=7cm,angle=0}
\end{center}
\caption{}
\label{diff}
\end{figure}

\end{document}